\documentclass[12pt]{iopart}

\usepackage{graphicx}
\usepackage{epsfig}
\usepackage{bm}

\def\be{\begin{equation}}
\def\ee{\end{equation}}
\def\bee{\begin{eqnarray}}
\def\eee{\end{eqnarray}}

\begin{document}

\title[Non-linear interaction of the tearing mode and electromagnetic turbulence.]{On seed island generation and the non-linear self-consistent interaction of the tearing mode with electromagnetic gyro-kinetic turbulence.}
\author{W A Hornsby$^{1,2}$, P Migliano$^1$, R Buchholz$^1$, D Zarzoso$^2$, F J Casson$^3$, E Poli$^2$, A G Peeters$^1$}
\address{$^1$ Theoretical Physics V, Dept. of Physics, Universitaet Bayreuth, Bayreuth, Germany, D-95447}
\address{$^2$ Max-Planck-Institut f\" ur Plasmaphysik, Boltzmannstrasse 2, D-85748}
\address{$^3$ CCFE, Culham Science Centre, Abingdon, Oxon., OX14 3DB, UK}

\ead{william.hornsby@ipp.mpg.de}

\date{\today}

\begin{abstract}

The multi-scale interaction of self-consistently driven magnetic islands with electromagnetic turbulence is studied within the three dimensional, toroidal gyro-kinetic framework.  It can be seen that, even in the presence of electromagnetic turbulence the linear structure of the
mode is retained.  Turbulent fluctuations do not destroy
the growing island early in its development, which then maintains a coherent form as it grows. 

The island is seeded by the electromagnetic turbulence fluctuations, which 
provide an initial island structure through nonlinear interactions and which grows at a rate significantly faster than the linear tearing growth rate.  These island structures saturate at a width that is approximately $\rho_{i}$ in size.   
In the presence of turbulence the island then grows at the linear rate even though the island is significantly wider than the resonant layer width, a regime where the island is expected to grow at a significantly reduced non-linear rate.   

A large degree of stochastisation around the separatrix, and an almost complete break down of the X-point is seen.  This significantly reduces the effective island width.  

\end{abstract}


Magnetic islands in a tokamak can lead to loss of confinement or even major disruptions of the plasma.  The tearing mode \cite{FUR63,FUR73}, specifically the neoclassical tearing mode (NTM) \cite{CAR86} is expected to set the beta limit in a reactor \cite{SAU97,WAE09}.  The dynamics of magnetic islands is the result of the interplay of variety of processes.  
On one hand, their instability is widely believed to be due to neoclassical effects, first 
of all to the non-linear drive caused by the bootstrap-current perturbation due to the 
island itself and often the polarisation current is invoked as a mechanism for its threshold \cite{ConPol,WIL96}.  Within the singular layer, centred around rational surfaces, (Parallel wave-vector of the mode, $k_{||}=0$)  the  assumptions of ideal MHD break down and magnetic reconnection may take place.  For a collisionless mode the mechanism of reconnection is the electron inertia, where the singular layer width, and in turn the growth rate, are related closely to the electron skin depth \cite{HAZ75}. When collisions become significant, it is plasma resistivity which determines the singular layer width, producing a classical collisional tearing \cite{FUR63} or, more relevant for present day tokamaks, a semi-collisional mode \cite{DRA77,Fitz10,Ade91}.   In high temperature, weakly collisional, plasmas the tearing mode width evolves very slowly at the resistive time scale. 

Drift-wave turbulence is widely acknowledged to be the cause of
anomalous transport in confined plasmas.   Turbulence is, in turn, is regulated by zonal and meso-scale flows \cite{ter00}.  Plasma turbulence and tearing modes occupy disparate time and length scales, with 
turbulence occupying the micro-scale defined by the ion gyro-radius and drift frequency.  
Tearing modes occupy a significant fraction
of a toroidal turn, however, early in their evolution, islands can be very narrow and thus comparable
to turbulent length scales.  As such, their evolution can not be considered to be independent of the turbulence \cite{MIL09}, and has been shown to be influenced by turbulence in gyro-fluid simulations in toroidal geometry \cite{POL10,ish07}.

Due to the large separation in time-scales and complexity of the problem, analytical theory in this field is difficult, here we approach the problem using massively parallel, state-of-the-art kinetic simulation.   The gyro-kinetic framework of equations has been highly successful when applied to the study of drift waves, turbulence and impurity transport.  Recently,
gyro-kinetics has been used to study magnetic reconnection in a highly magnetised system, however, always concentrating on two dimensions \cite{ROG07,NUM09,Wan05}. 
It has also been successfully used to study the internal kink instability \cite{Mish}.

\begin{figure}
\centering
\includegraphics[width=8.5cm,clip]{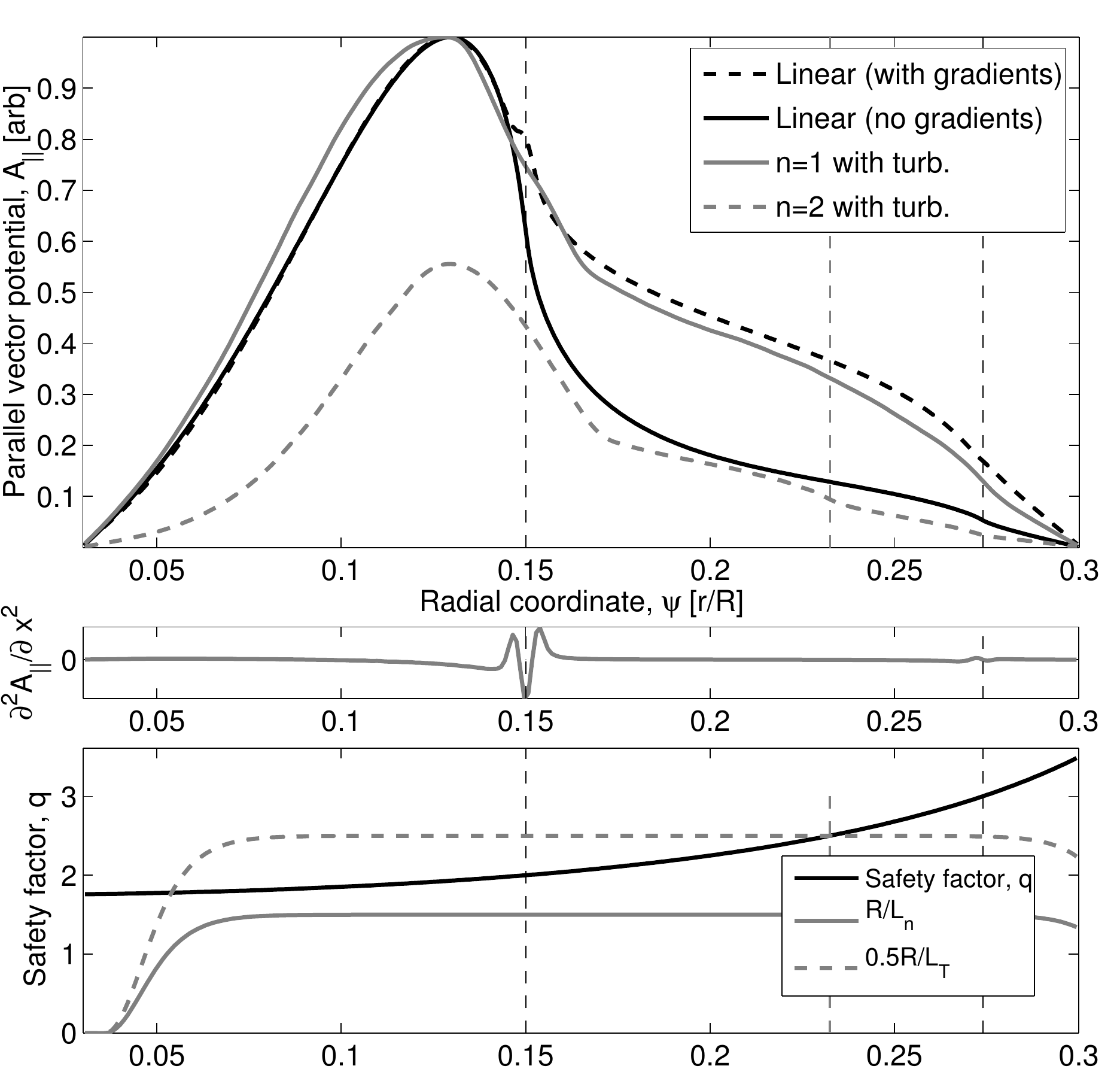}
\caption{(top) The radial profile of the parallel vector potential ($A_{||}$) for (black) for a linear calculation without the presence
of background density and temperature gradients, representative of the eigenfunction of the double tearing mode here with a $q=2$ and $q=3$ resonant layer.
(black, dashed line) Is the profile for a linear calculation with a background thermodynamic gradient $R/L_{n} = 1.5$, $R/L_{Te} = 5.0$. (Gray) is the profile in the
presence of electromagnetic turbulence once the island has been established.  The inlay shows the zoomed in electrostatic, $\phi$, profile from the linear calculations. 
 (bottom) The safety factor, q, profile utilised.  Vertical dashed lines represent the postions of the resonant layers.}
\label{Eigenfunction}
\end{figure}

The interaction of turbulence with static, imposed island structures has been previously investigated \cite{POL10,HorEPL,walwael,IshWael,ish12} and has uncovered aspects of multi-scale behaviour, producing
modified zonal flows and vortex structures \cite{HorVor}, and toroidal effects having a significant effect on temperature gradients through the island \cite{POL09,Hor10} and bootstrap current profiles \cite{HorVar}.  Studies utilising fluid models have uncovered complex non-linear interactions modified island growth rates \cite{Mur11}, even with simplified geometries.   Large structures produced by turbulence have been suggested as a possible mechanism for the generation of seed islands, vital to the triggering and evolution of the NTM \cite{itoh04,ish10}.  NTMs have been experimentally observed without the requirement of a seed-island trigger \cite{gude99}, and can also be triggered by an island generated by a classical tearing mode  \cite{reim02}.

Here we present gyro-kinetic calculations, which for the first time evolve a tearing mode in three dimensions with the inclusion of realistic toroidal geometry and plasma parameters where turbulence, zonal flows and magnetic island are allowed to evolve self-consistently.  This has required significant computational resources, with a single simulation representing approximately 0.5M hours of CPU time.  

\begin{figure}
\centering
\includegraphics[width=6.3cm,clip]{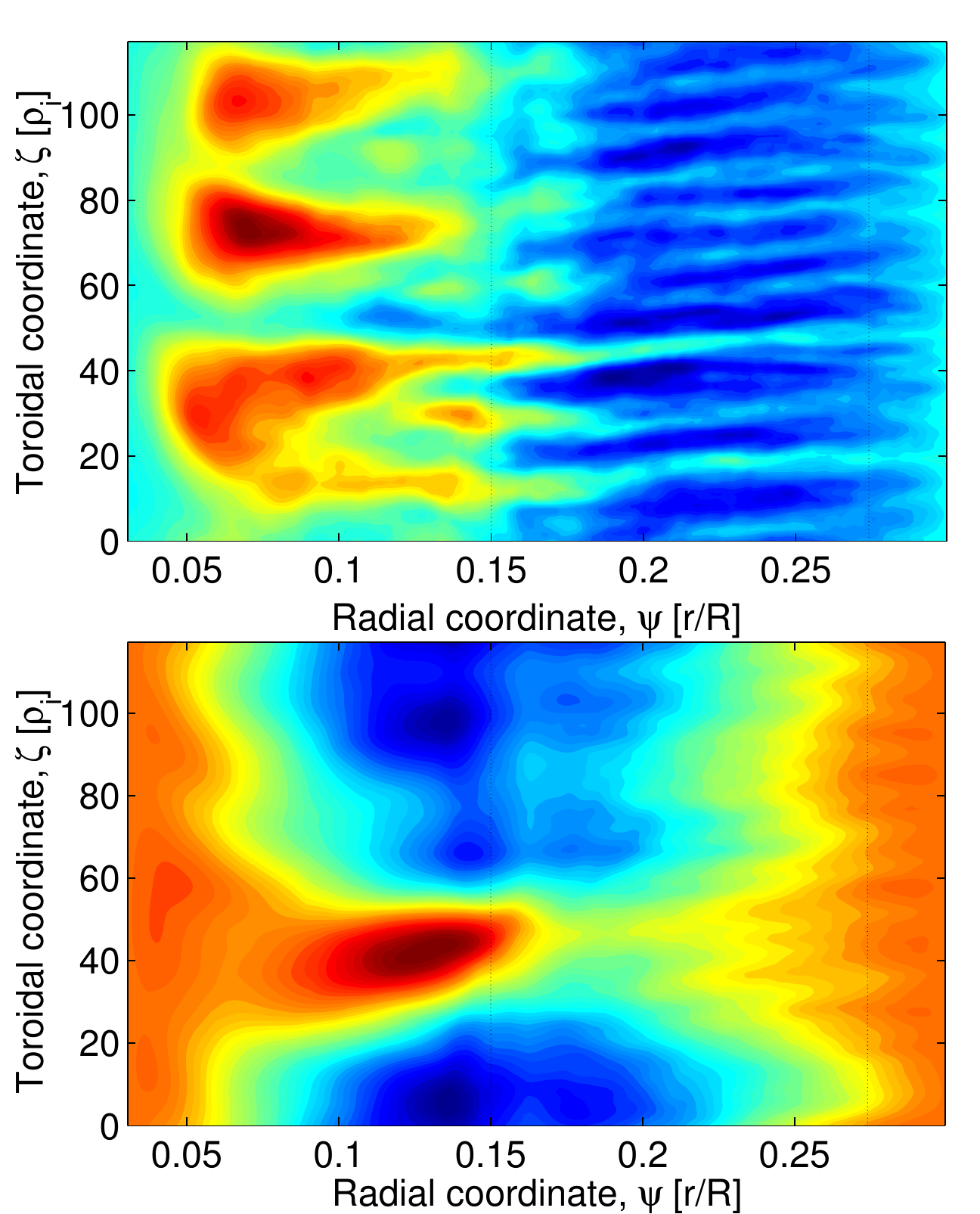}
\caption{(upper) A slice through the electrostatic potential and (lower) electromagnetic potential during a turbulence simulation at $t=165R/v_{th i}$ at a point
where the island structure has been established.  The vertical dashed lines represent the postitions of the resonant layers. }
\label{2dslices}
\end{figure}

The rest of this paper is organised as follows.  In Sections \ref{framework} and \ref{parameters} we outline the equations solved and the parameters and set up of the simulations.  In Sec.~\ref{seeisl} the generation of magnetic islands by electromagnetic turbulence is discussed, while in Sec.~\ref{magisl} the evolution of the magnetic island is analysed.   Conclusions are drawn in Sec.~\ref{conc}.

\section{Gyrokinetic framework}
\label{framework}

A self-consistent treatment of the tearing mode with turbulence requires a radial profile of geometric and thermodynamic quantities.  Used here is the global version of the gyro-kinetic code, GKW \cite{PEEGlo}.
Solved within GKW is the gyro-kinetic set of equations.  The full details can
be found in \cite{PEE09} and references found therein.  Here we outline the basic set of equations that are solved and in the 
next section the modification and assumptions used in driving the tearing instability.

The delta-$f$ approximation is used.  The final equation for the perturbed distribution function $f$, for each species, $s$,  can be written in the form 
\be 
{\partial g_{s} \over \partial t} + (v_\parallel {\bf b} + {\bf v}_D) \cdot \nabla f_{s} +  {\bf v}_\chi\cdot \nabla g_{s}  
-{\mu B \over m}{{\bf B}\cdot \nabla B \over B^2}{\partial f_{s} \over \partial v_\parallel} = S, 
\label{gyrovlas}
\ee
where $S$ is the source term which is determined by the background distribution function, $\mu$ is the magnetic moment, $v_{||}$ is the velocity along the magnetic field, $B$ is the magnetic field strength, m and Z are the particle mass and charge number respectively. Here,  $g = f + (Ze/T)v_{\parallel}\langle A_{\parallel} \rangle F_{M}$ is used to absorb the time derivative of the parallel vector potential $\partial A_{\parallel}/\partial t$ which enters the equations through Amp\`{e}res law.  
The background is assumed to be a shifted Maxwellian ($F_M$).   Here, $\rho_* = \rho_i / R$ is the normalised ion Larmor radius (where $\rho_i = m_i v_{th} / e B$ and $v_{th} = \sqrt{2 T_i / m_i}$).  Angled brackets denote gyro-averaged quantities.

The electrostatic potential is calculated from the gyro-kinetic quasineutrality condition, 
\bee 
\label{Poisson}
 \sum_{s}  Z_{s} e \int\langle f \rangle^{\dag}d^{3}{\bf v} +  \nonumber\\
 \frac{Z_{s}^{2}e^{2}}{T_{s}}\int (\langle\langle\phi\rangle\rangle^{\dag}-\phi)F_{Ms}({\bf v})d^{3}{\bf v} 
 = 0
\eee
where the first term represents the perturbed charge density and the second the polarisation density.  Similarly the parallel vector potential is calculated via Amperes law,
\bee
\nabla^2 A_\parallel + \sum_s {\mu_0 Z_s^2 e^2 \over T_s} \int {\rm d}^3 {\bf v} \, v_\parallel^2 \langle 
\langle A_\parallel \rangle \rangle^\dagger F_{Ms} = \nonumber\\
\sum_s {\mu_0 Z_s e \over T_s } \int {\rm d}^3 {\bf v} 
\, v_\parallel \langle g_s \rangle^\dagger 
\label{Ampere}
\eee

The gyro-average is then calculated as a numerical average over a ring with a fixed radius equal to the 
Larmor radius, 
$\langle G \rangle ({\bf X}) = {1 \over 2 \pi }\oint {\rm d} \alpha \, G({\bf X} + \boldsymbol{ \rho})$
where $\alpha$ is the gyro-angle.  The ${\dagger}$ denotes the complex conjugate operator.  This gyro-average is used in both the evolution equation of the 
distribution function, as well as in the Poisson and Ampere equations. The polarization in the latter equation 
is linearised (i.e. is calculated using the Maxwell background rather than the full distribution function). 

GKW uses straight field line Hamada \cite{HAM58} coordinates ($s,\zeta,\psi$) where $s$ is the coordinate along the magnetic field and $\zeta$ is the generalised toroidal angle. For circular concentric surfaces, the transformation of poloidal and toroidal angle to these coordinates is given by ($s,\zeta) = ((\theta + \epsilon \sin \theta)/ 2 \pi , [q \theta - \phi]/ 2 \pi)$.
The wave vector of the island is 
\begin{math}
k_\zeta^{I} \rho_i = 2 \pi n \rho_*,
\end{math}
where $n$ is the toroidal mode number.  GKW uses a Fourier representation in the toroidal direction. The radial direction is treated using finite-differencing with Dirichelet boundary conditions, so as to include geometry profile effects.  A Krook operator is used at the simulation boundaries to damp the perturbed distribution function so that a steady state is achieved. The operator limits the relaxation of the temperature profile, close to the simulations boundaries in a global run by damping the energy, but not the density or parallel velocity, in the perturbed distribution function.

\section{Parameters and set-up.}
\label{parameters}

The simulations presented utilise the parameters as follows and are summarised in Table \ref{params}.  The circular equilibrium geometry \cite{LAPcirc} is used.  Two values of $\beta_{e}$ are considered, $0.1\%$ and $0.2\%$ respectively.  The safety factor, $q$, at the at the plasma edge is set to $q_{a} = 3.5$.  As such, both $q=2$ and $q=3$ rational surfaces ($q=m/n$, where m is the poloidal
mode number) exist within our computational domain which extends from the plasma edge to 10\% of the radius from the magnetic axis.  

The normalised gyro-radius used is, $\rho_{*} = \rho_{i}/R =  0.005$.  This gives a $n=1$ toroidal mode number with normalised wave-number, $k_{\zeta}\rho_{i} =  0.053$.  Utilised here, in total, were 36 toroidal modes giving a maximum $k_{\zeta}\rho_{i} =  1.86$. The ion and electron temperatures are
assumed to be equal, $T_{i} = T_{e}$.  Resolutions in the parallel, parallel velocity, magnetic moment, radial directions are, $N_{s}=64, N_{v}=64, N_{\mu}=16, N_{x}=256$ respectively.  
This radial resolution places three radial grid points per singular layer width.  The linearly growing tearing mode is found to be a semi-collisional mode \cite{DRA77}.  As such, the singular layer width is closely related to the electron skin depth, which is given by,
\begin{equation}
\frac{\delta_{e}}{a} = \frac{\rho_{*}}{\sqrt{\frac{m_i}{m_e}}\sqrt{\beta_{e}}}\frac{R}{a}.
\end{equation}
which must be resolved.
The $\beta_{e}$ is the electron $\beta$ defined as $\beta_{e} = n_{e}T_{e}/(B_{0}^{2}/2\mu_{0})$,  where $n_{e}$ and $T_{e}$ are equilibrium density and
temperature and $B_{0}$ is the magnetic field strength at the outboard mid-plane magnetic axis. 

The tearing mode instability is driven by a non-homogeneous current density.  In our implementation this 
is introduced by applying an electron flow profile in the equilibrium.   The electron flow velocity is calculated self-consistently from the imposed $q$-profile  analogous to the method used in \cite{DRA77} for a kinetic calculation in slab geometry. 

Assuming that the background current is carried by the electrons, $J = -n_{e}v_{e}e$,  the current gradient is 
related to the flow gradient, $\partial u_{e}/\partial\psi$.  We neglect the gradient in the electron density and take the background electron flow velocity to be a flux function and ordering the electron flow velocity, $v_{e}$ to be significantly smaller than the electron thermal velocity.
For the current profile in this paper we use the same model as used by Wesson et. al. \cite{Wess,Has77}, where the current density profile is defined as, $j=j_{0}(1+(r/a)^{2})^{\nu}$, which introduces an electron flow which enters the equations via the source term, 
\begin{equation}
\nabla F_{Me} = -2 \frac{v_{||}}{v_{th}^{2}} \nabla u_{e} F_{Me}.
\end{equation}
The safety factor profile is calculated from the current profile, which has the analytic form,
\begin{equation}
q = q_{a}\frac{B_\phi}{R}\frac{r^{2}/a^{2}}{1 - (1 - r^{2}/a^{2})^{\nu+1}}
\end{equation}
where $q_{a}$ is the value of the safety factor at radial co-ordinate $\psi = a/R$.  An example of the safety-factor used can be seen in the bottom panel of Fig.~\ref{Eigenfunction}.  This is related to the $q$ on the axis
by $q_{a}/q_{0} = \nu + 1$, $\nu$ being an integer that determines that peaking of the current
gradient.

\begin{table}
\begin{center}
\begin{tabular}{| l | l |}
\hline
  \multicolumn{2}{|c|}{Simulation parameters.} \\
\hline
Aspect ratio, {R}/{a} & 3 \\
Electron beta, $\beta_{e}$ & $10^{-3} (0.1\%)$ \\
Current peaking parameter, $\nu$ & 1 \\
Mass ratio, $m_{i}/m_{e}$ & 1836 \\
$q$ at edge, $q_{a}$ & 3.5 \\
$q$ on axis, $q_{0}$ & 1.75 \\
Smallest mode number, $k_{\zeta}\rho_{i}$  & 0.053 \\
Largest mode number, $k_{\zeta}\rho_{i}$  & 1.86 \\
Number of toroidal modes, N & 36 \\
$\rho_{*} = \rho_{i}/R$ & 0.005 \\
$T_{i} = T_{e}$ & \\
$\max(v_{||}/\mu)$ & 4/8\\
\hline
\end{tabular}
\caption{A summary of the simulation parameters used.}
\label{params}
\end{center}
\end{table}

The current peaking parameter utilised in the work here is, $\nu=1$, which is found to provide a linearly unstable semi-collisional tearing mode \cite{HorLin14}, with a positive stability parameter, \cite{FUR63} defined as, 
\begin{equation}
\left.\Delta' = \frac{1}{A_{||}}\frac{\partial A_{||}}{\partial r}\right|^{r_{s}^{+}}_{r_{s}^{-}}
\end{equation}
 across the singular layer, whose position is denoted by $r_{s}$ and upper and lower boundaries by $r_{s}^{+}$ and $r_{s}^{-}$ .  This parameter represents the discontinuity in $\partial A_{||}/\partial r$ and measures whether mode growth is energetically favourable ($\Delta' > 0$ for the mode to grow).
The tearing mode has the familiar \cite{NISH98} linear structure of which is shown in Fig.~\ref{Eigenfunction}, showing the position of the resistive layers at the $q=2$, $q=2.5$ and $q=3$ rational surfaces which are within the domain.  Tearing modes at higher toroidal mode numbers were found to be linearly stable.

The normalised collision frequency (to the trapping/de-trapping rate), $\nu_{*} = 4\nu_{ei}/3\sqrt(\pi\epsilon^{3}) = 0.12$, where  the collision frequency is defined as
$\nu_{ei} = \frac{n_{i}e^{4}\log{\Lambda_{ei}}}{4\pi\epsilon_{0}^{2}m_{e}^{2}v_{e}^{3}}$.  In this paper we only consider pitch-angle scattering of the electrons from the ions and neglect the neoclassical terms, as such the NTM is not explicitly considered.

The density and temperature profiles have the radial form,
\begin{eqnarray}
\frac{\partial n_{s}}{\partial \psi} = \frac{1}{2}\frac{R}{L_{ns}}(\tanh{(x-x_{0}+\Delta x)}-\tanh{(x-x_{0}-\Delta x)})\nonumber\\
\frac{\partial T_{s}}{\partial \psi} = \frac{1}{2}\frac{R}{L_{Ts}}(\tanh{(x-x_{0}+\Delta x)}-\tanh{(x-x_{0}-\Delta x)})\nonumber
\end{eqnarray}
where $R/L_{n} = -(R/n)\partial n/\partial\psi$ and $R/L_{T} = -(R/T)\partial T/\partial\psi$.  An example of the radial profiles used is shown in the middle panel of Fig.~\ref{Eigenfunction}.  The gradients used in the rest of this paper are $R/L_{n} = 1.5$ and $R/L_{T} = 5.0$ at the reference radius, $x_{0} = r/a = 0.5$ for both ions and electrons, which are large enough to drive micro-instabilities which lead to turbulence, a snapshot of which is shown in Fig.~\ref{2dslices}, but not large enough to completely stabilise the linear tearing mode \cite{Dra83}.  This was confirmed by linear simulations.  In all cases the simulation was initialised with noise.

\section{Seed island generation.}
\label{seeisl}

Large structures produced by turbulence have been suggested as a possible mechanism for the generation of seed islands, vital to the triggering and evolution of the NTM.  In this section the early time evolution of the simulation is studied, corresponding to the time range $t=0$ to $t=90 R/v_{th i}$ in Fig.~\ref{Islandtrace} .  

\begin{figure}
\centering
\includegraphics[width=8.0cm,clip]{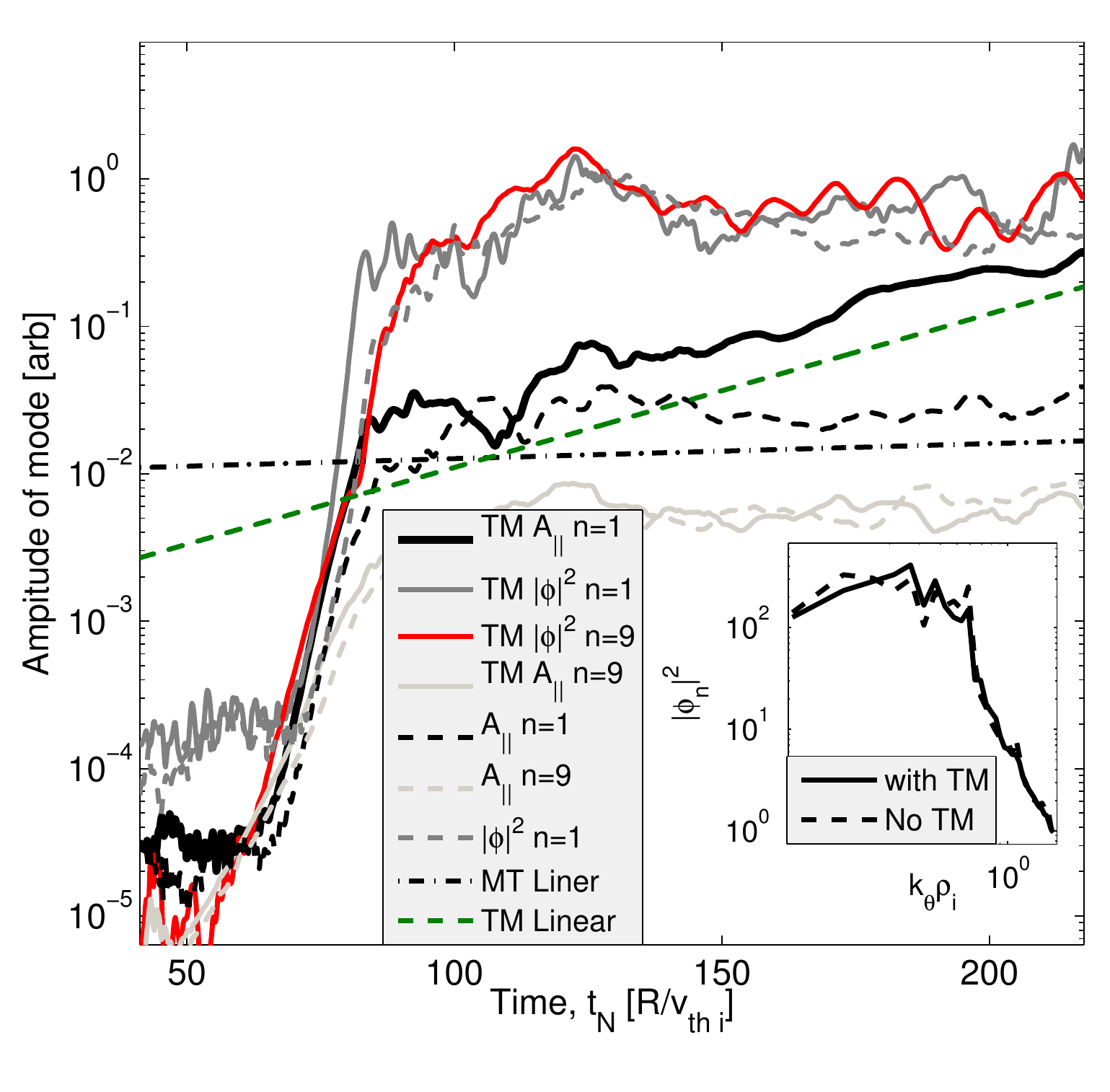}
\caption{Time traces of the squared electrostatic potential ($|\phi|^{2}$) (Grey lines) and electromagnetic potential ($A_{||}$) (Black lines) for a simulation with an imposed current profile (Solid lines) and without (Dashed lines).  $\phi$ is largely unaffected, maintaining a statistical steady-state.  $A_{||}$, however, continues to grow once a seed island structure is established at approximately $t=90 R/v_{th i}$.  The (red) curve is the electrostatic potential time trace of the $n=9$ toroidal mode.  The linear scalings for the microtearing (black dot-dashes) and linear tearing mode (green dashes) are also plotted.  In the inlay the time averaged electrostatic fluctuation spectrum of the case without current drive is shown.  }
\label{Islandtrace2}
\end{figure}

In Fig.~\ref{Islandtrace2} time trace of the $n=1$ and $n=9$ toroidal mode amplitude for both the electrostatic potential and the electromagnetic potential is plotted for two sets of simulations,  one with and the other without an imposed equilibrium electron current gradient.

\begin{figure}
\centering
\includegraphics[width=8.0cm,clip]{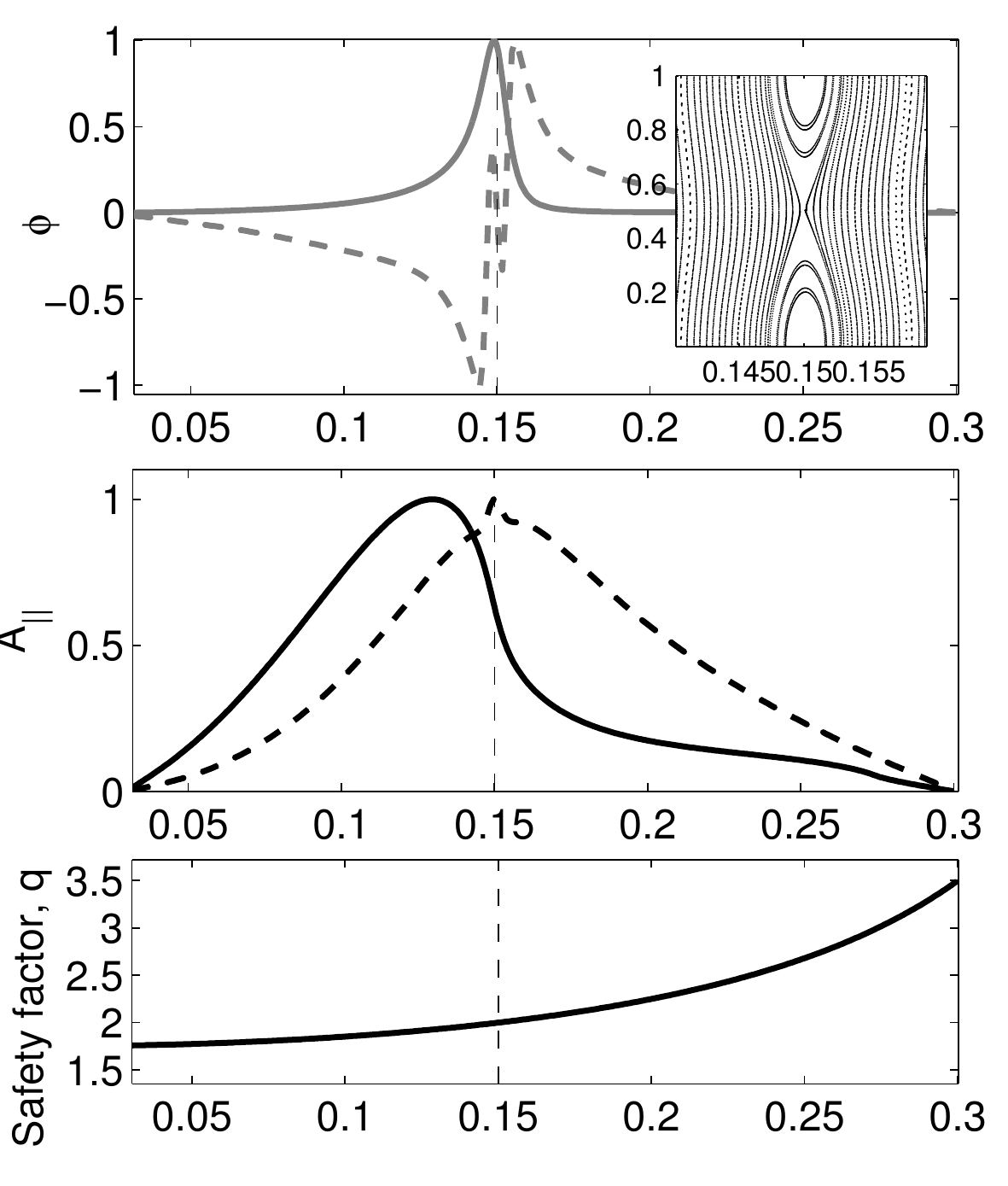}
\caption{(top) The radial profile of the electrostatic potential for  a tearing mode (solid) micro-tearing mode (dashed) in the $n=1$ toroidal mode, while (middle) shows the parallel vector potential profile and (bottom) the corresponding safety factor profile.  Inlaid in the top is the poincare plot showing the magnetic island produced by a micro-tearing mode at the $q=2$ rational surface. }
\label{MTstruct}
\end{figure}

We see in each of these, in the initial phase of the simulation, an approximately exponential growth of the mode amplitude before saturating at approximately 100$R/v_{th i}$ and progression towards a statistical steady state.    The mode amplitudes reach saturation very rapidly with respect to an island evolving without turbulence,  the turbulence providing a large seed island structure.  This seed island is produced by non-linear interactions.  When the electromagnetic turbulence is resolved, a perpendicular slice of which is shown in  Fig.~\ref{2dslices}, the tearing mode displays a much more rapid growth in the initial phase compared with linear tearing mode theory and the linear growth of electromagnetic micro-instabilities.  

In linear simulations without the tearing drive, it is found that a global micro-tearing mode \cite{appMT} is also unstable at this toroidal mode number with a growth rate, $\gamma=0.00236 v_{th i}/R$ and mode frequency, $\omega=-0.212 v_{th i}/R$, the negative sign denoting the electron diamagnetic direction.  The radial Eigenfunctions of the of the electric and magnetic potential for the micro-tearing mode are shown in Fig.~\ref{MTstruct} (dashed lines) as well as the Eigenfunctions of the tearing mode when the background current is kept (solid lines).  It can be seen in inlay that this mode can produce magnetic islands at the $q=2$ rational surface even though the mode has a different structure to a tearing mode at the same rational surface.  The growth of the mode embedded in turbulence, is significantly more rapid than the growth rate of the $n=1$  micro-tearing mode found when the background current gradient is neglected.  The time scaling of the latter is shown as a dot-dashed line in the main panel of Fig.~\ref{Islandtrace2} and the scaling of the linear tearing mode growth rate as the green dashed line.  

In this phase it can be seen that the island mode amplitude ($n=1$, $k_{\theta}\rho_{i}=0.055$) closely traces the amplitude of the turbulence ($A_{||}\propto |\phi|^{2}$) with $|\phi|^{2}$ taken from the peak of the turbulence intensity spectrum which has the wave-number, $k_{\theta}\rho_{i}=0.35$ (the trace is plotted in Fig.~\ref{Islandtrace2} (red line)).  An example of the turbulence spectrum is plotted in the inlay of Fig.~\ref{Islandtrace2} which has been time averaged from $t=100 R/v_{th i}$ to $t=250 R/v_{th i}$. 
Both these traces are plotted in the main panel of Fig.~\ref{Islandtrace2}, showing that the early time growth (between $t=60 R/v_{th i}$ and $t=90 R/v_{th i}$) of the island mode is via non-linear coupling with the electromagnetic turbulence.   Without turbulence the island evolves linearly (See inlay of Fig.~\ref{Islandtrace}) with a growth rate of $\gamma=0.024 v_{th i}/R$.

\begin{figure*}
\centering
\includegraphics[width=16.5cm,clip]{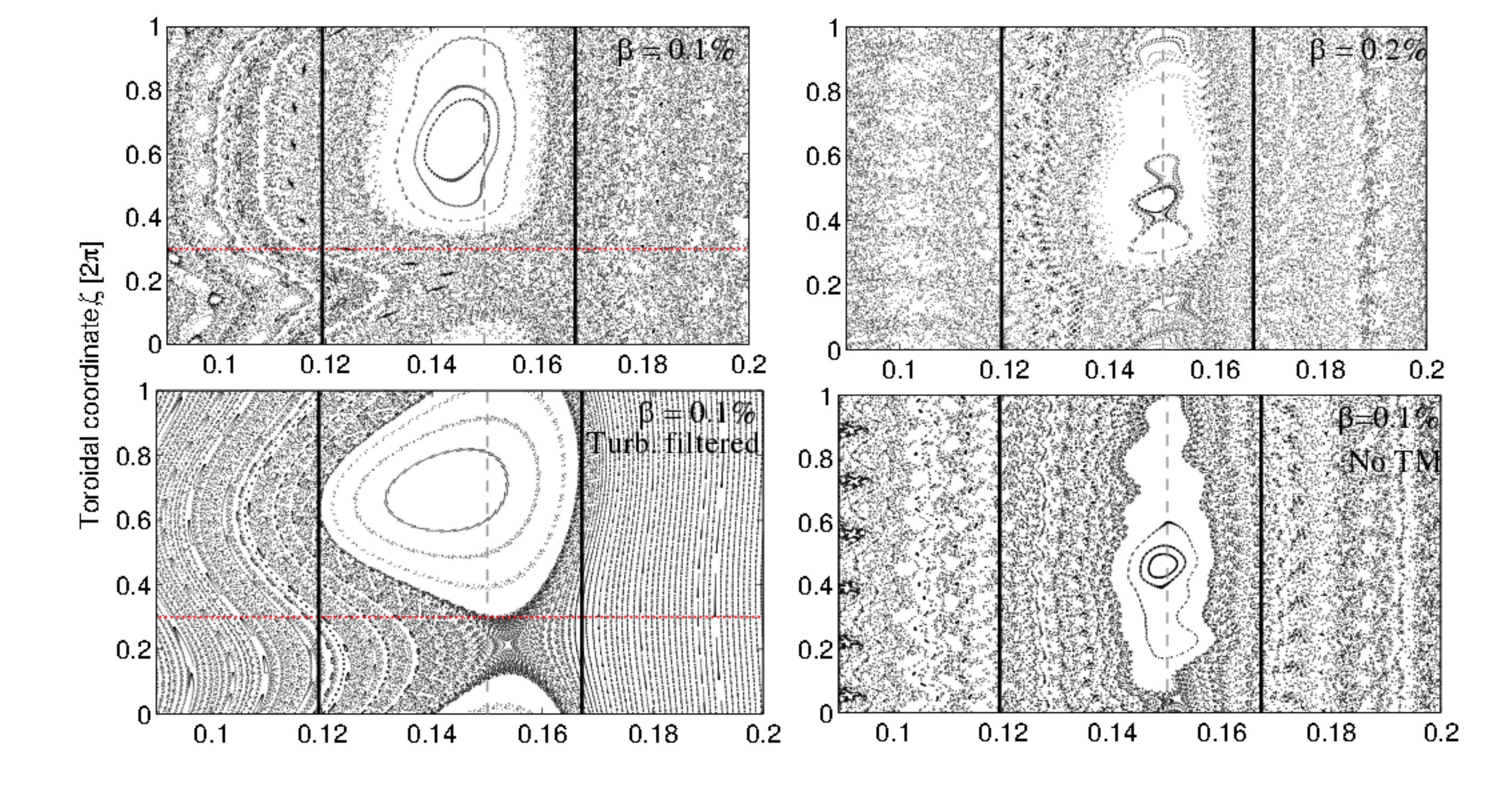}
\caption{Poincar\'{e} plot of the magnetic field lines as they pass through the low field side ($s=0$).  (Top left) The field lines
trace the modified form of a magnetic island as it evolves in electromagnetic turbulence, $\beta = 0.1\%$ and  (Top right) the equivalent plot for $\beta=0.2\%$.  Vertical dashed lines denote the position
of the rational surface and the red dots represent the original 
positions of the field lines. Solid lines represent the positions of the island seperatrix when turbulence is neglected. (Bottom left) Island in turbulence with $\beta = 0.001$ but all the modes apart from $n=1$ filtered out.  (Bottom right) Island structure generated by EM turbulence when no current drive is present at approximately, $t=80 R/v_{th i}$ (i.e. there is no linear tearing mode driven).}
\label{PoincareT}
\end{figure*}

The non-linear growth the the MT mode produces $A_{||}$ structures approximately $\rho_{i}$ in size, larger than the singular layer width of linear tearing mode theory theory.  Therefore, in the presence of turbulence linear tearing mode stability
is irrelevant since the turbulence, even at an electron beta $\beta_e \sim 0.1\%$, produces an island size for which linear theory is no longer, in principle, 
applicable.  This is further discussed in the following section.  The seed island structure generated by these interactions can be seen in
the bottom right panel of Fig.~\ref{PoincareT}, which shows a coherent island structure at the rational surfaces even without the presence of tearing mode drive, however, in this case the island saturates and remains small in size.
 
\section{Magnetic island evolution.}
\label{magisl}

\begin{figure*}
\centering
\includegraphics[width=11.5cm,clip]{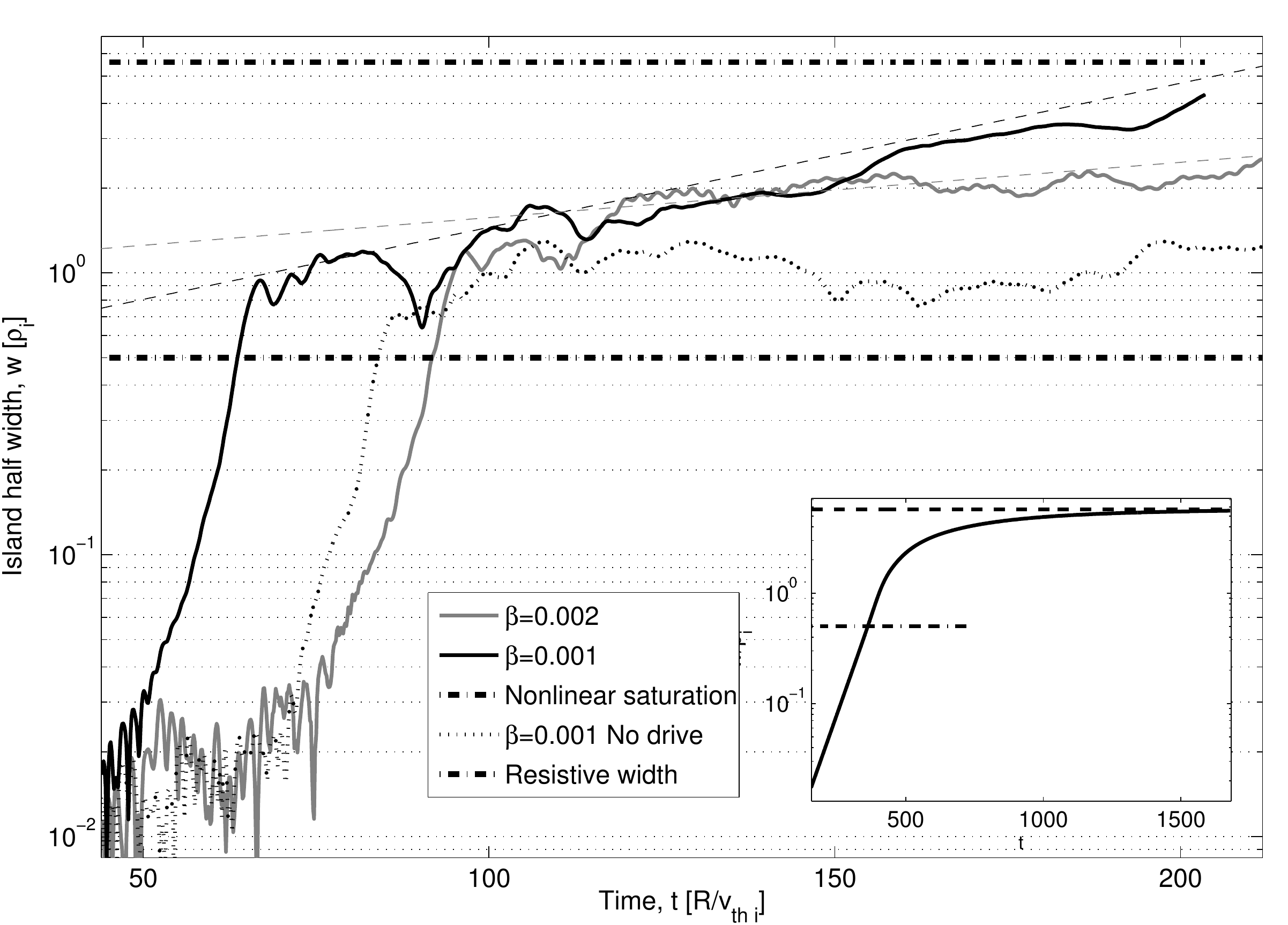}
\caption{The time traces of the width of the $m=2$, $n=1$ magnetic island (in units of $\rho_{i}$) as a function 
of time in the presence of electromagnetic turbulence.  Plotted is the trace for three simulations, one without the tearing mode drive, and two different values of $\beta_{e}$.  The inlay shows the island width for a non-linear simulation
without turbulence, showing the island evolve through the linear phase, non-linear phase toward saturation with the same parameters.  The saturated width and resonant layer widths are plotted in both.}
\label{Islandtrace}
\end{figure*}

In this section we will concentrate on the evolution of the island after $t=90 R/v_{th i}$.  Once the seed island is established the mode amplitude continues to grow, corresponding to a growing magnetic island.  At this point in the simulation it can be seen from the main panel in Fig.\ref{Islandtrace}, while the  electrostatic potential mode amplitude reaches a saturated state, the parallel vector potential, $A_{||}$, continues to grow in amplitude, representing a growing magnetic island. 

Comparing the traces with (solid curve) and without (dashed curve) the tearing mode drive due to the radial gradient of the background
current, one observes that in the latter case the parallel vector potential ($A_{||}$) mode amplitude saturates after the initial phase, 
while in the former case it grows in amplitude throughout the simulation, developing increasing tearing parity and generating a growing magnetic island.    The island width is calculated from $A_{||}$ using the expression,
\begin{equation}
w = \sqrt{4q A_{||}/B_{t}\hat{s}}.
\end{equation}

From Fig.~\ref{Eigenfunction} which is taken from a simulation when the drive is present, we see the distinctive \cite{NISH98} radial eigenfunction of a $m=2$, $n=1$ tearing mode.  
Therefore, even though electromagnetic turbulence is present the magnetic island continues to grow and maintain a coherent structure.  Turbulence does not disrupt the growth of the tearing mode, even for island sizes of the order of the ion Larmor radius, where the island width is comparable to the length scales of the turbulent eddies.

As a comparison, the inlay in right panel of Fig.~\ref{Islandtrace} shows the island when the turbulence is unresolved, this is achieved by running a nonlinear simulation using the same parameters but with only the $n=1$ and $n=0$ toroidal modes kept.  In this case there is no small scale turbulence and initially, the mode grows exponentially  in a linear phase until the island size reaches the plotted singular layer width (The singular layer width is plotted in the inlay as well as in the main Fig~\ref{Islandtrace} as the black dash-dotted line).  The island then enters the Rutherford non-linear phase at approximately 270 $R/v_{th i}$ before
approaching saturation.     At this island size the mode enters the expected Rutherford \cite{RUTH73} non-linear regime and grows algebraically ($w \propto t$) until it finally saturates.  The saturated island half width is $w = 5.7\rho_i$ at $\beta_{e}=0.1\%$, which is also indicated by the dashed horizontal lines in both the inlay and the main figure.  

The three curves in the main panel of Fig.~\ref{Islandtrace} are the island widths when the full range of toroidal modes are kept, and thus the tearing mode interacts with electromagnetic turbulence.   One curve with and one without the current drive at $\beta_{e}=0.1\%$ and one with the current drive at $\beta_{e}=0.2\%$. 
It is evident that once the island is established, it evolves at a rate closer to the standard linear growth rate, whose corresponding scaling is denoted by the dashed lines.  These linear growth rates were numerically calculated to be $\gamma=0.024 v_{th i}/R$ for $\beta_{e}=0.1\%$, and $\gamma=0.0090 v_{th i}/R$ for $\beta_{e}=0.2\%$.  Without turbulence present the island evolves at these growth rates within their respective linear regimes.  However, the islands here are much larger than the singular layer \cite{DRA77} width, denoted by the horizontal dashed line and calculated from the case without turbulence.  The singular layer width is the maximum island size at which linear theory is valid, and islands larger than this would be expected to  evolve at the much slower, non-linear rate.  From the radial profiles of $A_{||}$ in Fig.~\ref{Eigenfunction} we see that
the profile from a point in a turbulence simulation is almost unchanged from its linear Eigenfunction, however, the fact that the island grows linearly at this amplitude implies that the turbulence is disrupting the Rutherford non-linear mechanism \cite{RUTH73} or its threshold.  It is therefore seen that even though strong electromagnetic turbulence is present the magnetic island continues to grow and maintain a coherent structure.  

Apart from the small scale structures that influence the island in a turbulent
setting, also large scales develop.  Higher mode numbers, for 
example the $m=4$, $n=2$ mode whose eigenfunction (and corresponding resonant layers) is also shown in the dashed grey line 
in Fig.~\ref{Eigenfunction}, have a significant amplitude with respect to the $n=1$ mode and modify the island structure significantly. 
These higher mode number tearing modes, as well as drift waves at this mode number are linearly stable and are generated here by non-linear 
interactions.   Finally there is a slight shift in its radial position  of the island with respect to the $q=2$ 
rational surface of the background magnetic field (the latter is denoted by the vertical dashed black line).  This shift is caused by a finite amplitude 
of the vector potential in the $n=0$ toroidal mode.

Comparing the Poincar\'{e} plots of the island, with and without turbulence, the shape of the magnetic island in a turbulent setting is modified as can be seen in Fig.~\ref{PoincareT} which 
shows a Poincar\'{e} plot of the magnetic field lines through the low field side.  The radial width of the closed island structure is significantly smaller when the small
scale modes are retained.  The magnetic island, when the turbulence is filtered out is shown in the top right panel.  The black vertical lines in the other panels are the positions of the seperatrix as calculated from this island.

Even for the small $\beta_{e}$ that is used here, the island is 
approximately $30\%$ smaller, with the region around the separatrix and X-point highly stochastised.  The degree of stochastisation which
increases the higher the $\beta_{e}$.  Plotted in the top left panel is the island with $\beta_{e}=0.2\%$ with approximately the same mode amplitude and it is evident that the stochastisation, and effective reduction in island width is greater. 

A slight shift in the islands radial position is seen, with respect to the $q=2$ rational surface as denoted by the vertical dashed black line.  This shift is caused by a finite amplitude 
of the vector potential in the zero toroidal mode modifying the equilibrium current profile (the original position is returned if we filter out the zonal mode).  We see from this figure also that the presence of higher mode numbers, for example the $m=4$, $n=2$ mode whose Eigenfunction (and corresponding resonant layers) is also shown in the dashed grey line in Fig.~\ref{Eigenfunction}, modifies the island structure significantly.   These higher modes are linearly stable to the tearing mode, however, here they are generated by non-linear interactions.

\section{Conclusions.}
\label{conc}

The self consistent interaction of a linearly unstable tearing mode with gyro-kinetic electromagnetic turbulence was investigated in three dimensional toroidal geometry.  The main results can be summarised as follows:

\begin{itemize}

\item Islands as long wavelength coherent structures are sustained in electromagnetic turbulence.  The turbulence does not destroy the magnetic island early in its evolution, when the radial scale length is comparable to the turbulence scale length.  In fact, non-linear interactions pump energy into the island mode making it grow to a large amplitude rapidly. 

\item The turbulence, however, causes a stochastisation around the island seperatrix and X-point making it difficult to define the boundaries of the magnetic island and reducing its effective size.  This reduction is proportional to the plasma $\beta$, with a larger $\beta$ having a larger effect.  This has profound implications for boundary layer theories of island stability where a distinct separation between the plasma within and outside the magnetic island is required.

\item The turbulence provides a seed structure that is approximately an ion-gyroradius in size, that is large enough to generate an island that is 
larger than the resistive layer width.   Therefore the island is directly evolved into a regime which should be non-linear, however we see the island evolving at a rate close to the linear calculation implying a disruption of the Rutherford non-linear growth mechanism.

\item A zero toroidal moment component of the parallel vector potential causes a modification of
equilibrium current profile that modifies the position of the rational surfaces and causes a shift in the
position of the magnetic island.

\end{itemize}

Stochastisation of the magnetic island is expected to have profound effects on the boundary layer between it and the bulk plasma as well as have a significant role in the particle and heat transport.
The polarization current stabilization \cite{WIL96}  connected with the rotation of the island that critically depends on the boundary
layer can be expected to be strongly modified.

The mode frequency, and in particular, its direction is vital in predicting the non-linear stability of the NTM.  Specifically the sign of the mode frequency determines whether the polarisation current has a stabilising or destabilising effect \cite{ConPol,Smol93} and 
to which extent it is screened by other effects in toroidal geometry \cite{Pol05}. The rotation is 
also known to affect the the degree of density profile flattening within a small island \cite{Sicc11, Dzar14} 
which, in turn, will have consequences on the strength of the neoclassical drive \cite{Berg09}. An 
in-depth investigation of self-consistent island rotation is beyond the scope of this paper and is left for a future publication.

\ack

A part of this work was carried out using the HELIOS supercomputer system at Computational Simulation Centre of International Fusion Energy Research Centre (IFERC-CSC), Aomori, Japan, under the Broader Approach collaboration
between Euratom and Japan, implemented by Fusion for Energy and JAEA.

The authors would like to thank the Lorentz Centre, Leiden for their support.  Useful and productive conversations with
the attendees of the meeting, ``Modelling Kinetic Aspects of Global MHD Modes'' are gratefully acknowledged.

\end{document}